\documentclass[12pt]{iopart}
\usepackage{iopams}
\usepackage[abbr]{harvard}
\usepackage[dvips]{graphicx}
\begin{document}

\title[p-type TCOs based on Zinc Oxide Spinels]{p-Type Zinc Oxide Spinels: Application to Transparent Conductors and Spintronics}

\author{Maria Stoica and Cynthia S Lo}

\address{Department of Energy, Environmental and Chemical Engineering, Washington University, 1 Brookings Drive, St. Louis, MO 63130-4899, USA}
\ead{clo@wustl.edu}
\begin{abstract}
We report on the electronic and optical properties of two theoretically predicted stable spinel compounds of the form ZnB$_2$O$_4$, where B = Ni or Cu; neither compound has been previously synthesized, so we compare them to the previously studied p-type ZnCo$_2$O$_4$ spinel. These new materials exhibit spin polarization that is ideal for spintronics applications, and broad conductivity maxima near the valence band edge that facilitate p-type dopability. We show that $3d$ electrons on the octahedrally coordinated Zn atom fall deep within the valence band and do not contribute significantly to the electronic structure of the material, while the O $2p$ and tetrahedrally coordinated B $3d$ electrons hybridize broadly in the shallow valence states, resulting in increasing curvature (i.e., decreased electron effective mass) of valence bands near the band edge. In particular, ZnCu$_2$O$_4$ exhibits high electrical conductivities near the valence band edge that, at $\sigma = 2 \times 10^{4} \ \textrm{S}/\textrm{cm}$, are twice the maximum found for ZnCo$_2$O$_4$, a previously synthesized compound in this class of materials.  This material also exhibits ferromagnetism in all of its most stable structures, which makes it a good candidate for further study as a dilute magnetic semiconductor.
\end{abstract}

\submitto{\NJP}
\maketitle

\section{Introduction}
Ternary oxides, particularly doped films and compounds of oxides of tin, indium, and zinc, have long been known to possess strong transparent conducting properties \cite{Chopra19831}. Depending on the components in the compounds, ternary oxides may also exhibit magnetic properties and find use as dilute magnetic semiconductors (DMSs) in spintronics \cite{Prinz27111998}. Many of the best-performing transparent conducting oxides (TCOs) available today are n-type, although there have been many attempts to produce p-type ZnO and CuAlO$_2$ films. Our focus is on ZnO films, as the ternary compounds of these oxides show promise for high electrical conductivities but are largely undiscovered. The oxygen vacancies and interstitial defects in ZnO have been shown to cause Fermi level pinning in these compounds, making it difficult to achieve p-type conductivities.  It is this lack of highly conductive p-type materials that presents an obstacle to attaining progress in transparent electronics and spintronics. Ternary oxides, however, may prove to be the gateway to better performing p-type materials. Although many ternary oxides are insulating, they may be doped to induce semiconducting behavior. There are 140 known oxide spinels, with possibly many more structures that have yet to be synthesized. A common means of searching for ternary oxides has been to sample the triangular phase space and determine whether the most favorable structures are thermodynamically stable \cite{Hautier2010,Hautier:2013fk}. 

Of these ternary oxides, the zinc oxide spinels (AB$_2$O$_4$, where A = Zn) are particularly attractive due to their low cost and unique doping characteristics \cite{zakutayev2012cation}. Specifically for p-type ZnCo$_2$O$_4$ and NiCo$_2$O$_4$, high conductivities in these materials are attributed to off-stoichiometric compositions, especially when synthesized in Zn/Ni rich environments. It has been shown that NiCo$_2$O$_4$ exhibits high conductivities, on the order of $\sigma = 10^2 \ \textrm{S}/\textrm{cm}$, in part due to self-doping through the preference of the Ni atom for the T$_d$ (tetrahedrally coordinated) site over the O$_h$ (octahedrally coordinated) site in the normal spinel structure. According to multiple theoretical investigations, ZnCo$_2$O$_4$ is a normal spinel at low temperatures. As temperature increases, so do the number of anti-site defects, and at temperatures above 800 K, the structure becomes heavily non-stoichiometric \cite{Paudel2011}. The anti-site defects create donor and acceptor states, and thus are favorable for improving the electrical conductivity of the material \cite{Paudel2011doping}. The compound is, however, more optically absorbent than generally desirable for transparent electronics. 

In this work, we look at yet-to-be synthesized first-row transition metal zinc oxide spinels, and compare their electrical transport and optical properties to those of the recently characterized ZnCo$_2$O$_4$ \cite{PhysRevB.84.205207}. There have been many promising results on the role of the Zn atom in these structures, and to complement these results, we would like to investigate the effect of exchanging the Co (B) atom for a higher valence atom.

\begin{figure}
\includegraphics[width=200pt]{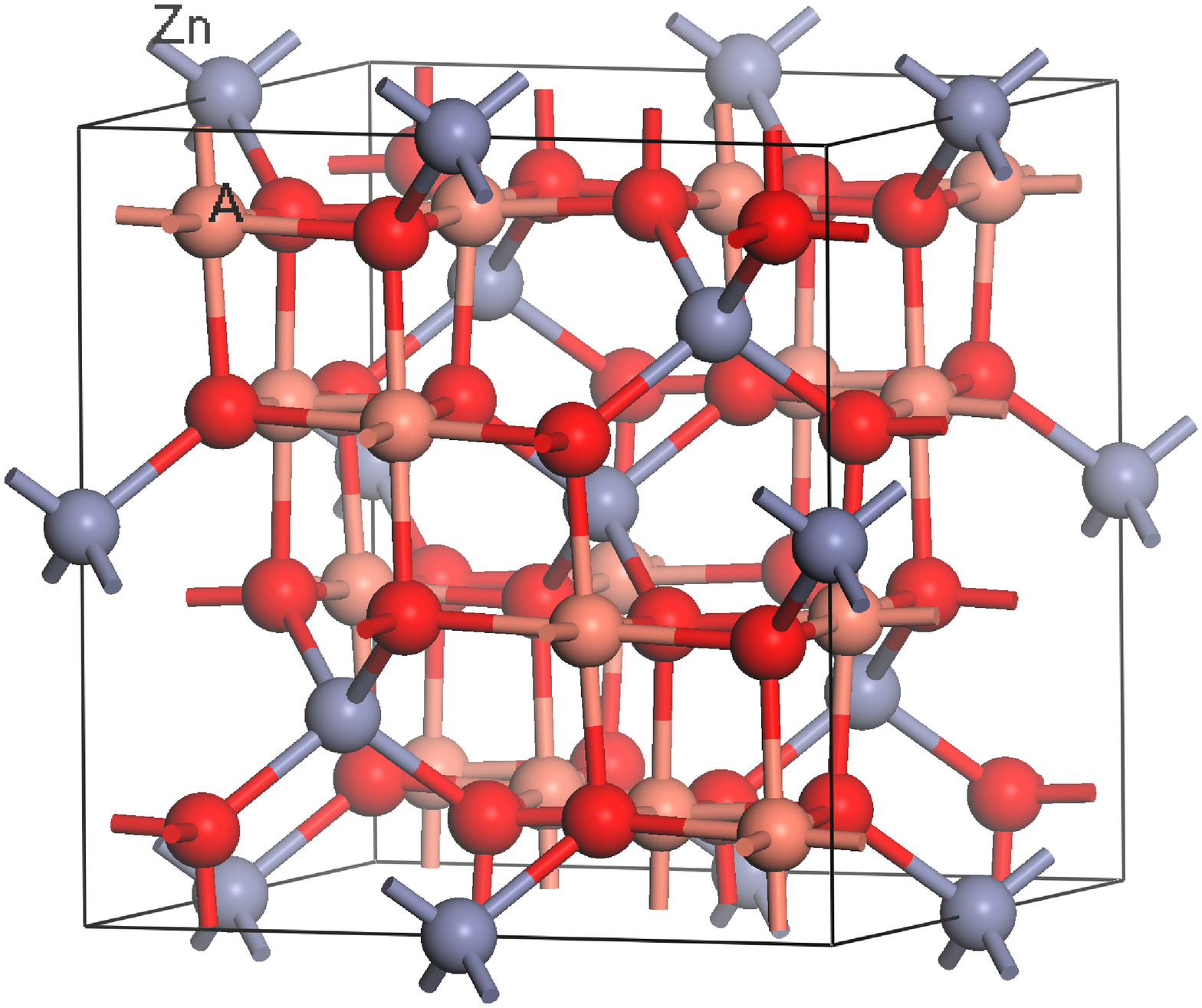}
\includegraphics[width=230pt,trim=0 300 0 0]{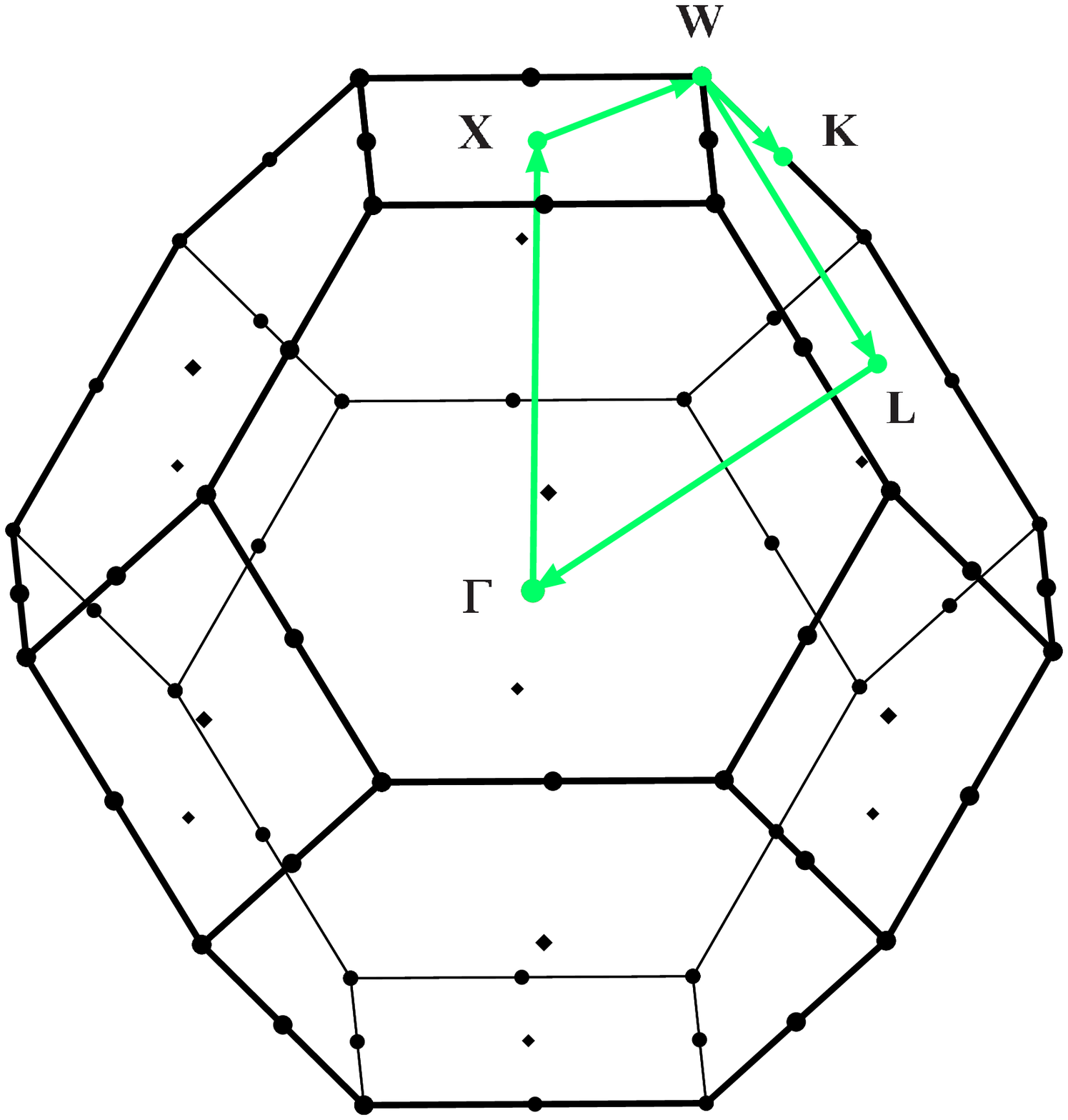}
\caption{Unit cell of the cubic spinel ZnA$_2$O$_4$ (left), and the high symmetry path for the corresponding first Brillouin zone (right). \label{cubicstructure}}
\end{figure}

\begin{figure}
\includegraphics[width=200pt]{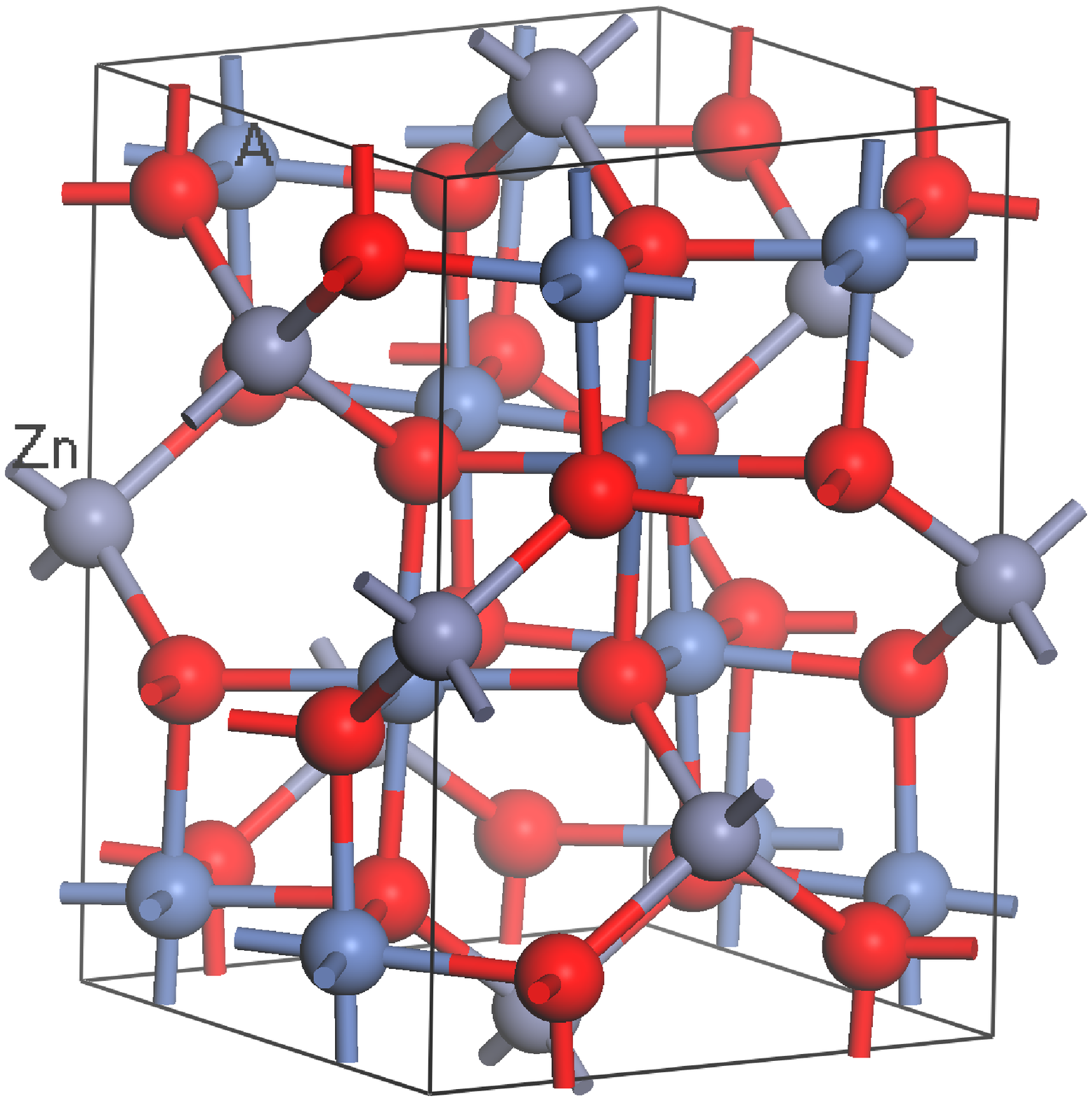}
\includegraphics[width=230pt,trim=0 300 0 0]{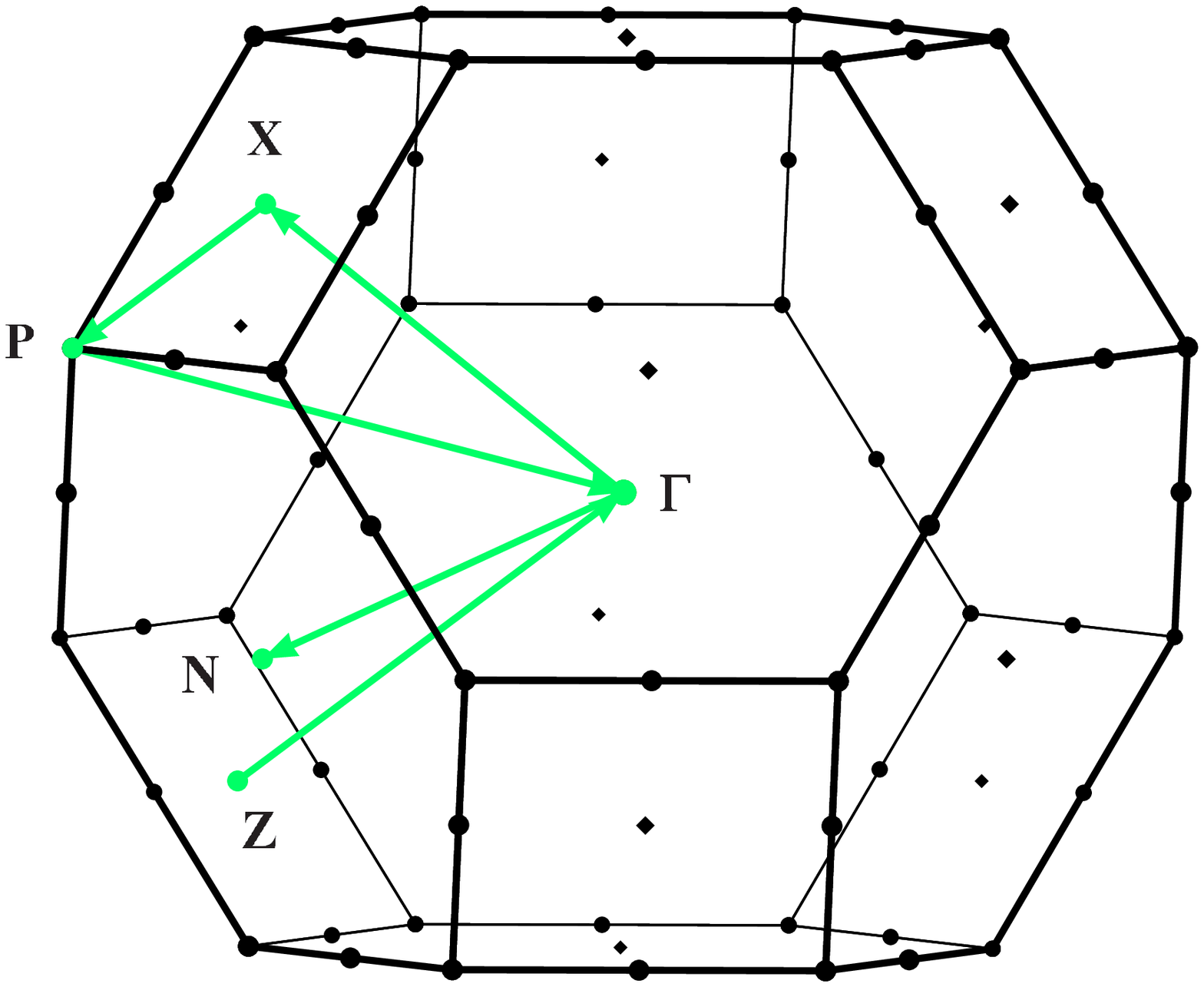}
\caption{Unit cell of the tetragonal spinel ZnA$_2$O$_4$ (left), and the high symmetry path for the corresponding first Brillouin zone (right). \label{tetragonalstructure}}
\end{figure}

The structures of the zinc oxide spinels chosen for this study are based on recent developments in accelerated materials screening and discovery. \citeasnoun{Hautier2010} developed an algorithm, which combines machine learning techniques and high throughput \textit{ab initio} calculations, to build a model for predicting the structure of known ternary oxide compounds and finding new, stable compounds. Among the 209 compounds predicted by their algorithm are two first row transition metal zinc oxide spinels -- ZnNi$_2$O$_4$ and ZnCu$_2$O$_4$. Spinels may have different configurations, of which the most common is the normal cubic (Fd$\bar{3}$m). The distorted tetragonal (I4$_1$/amd) structure is another common structure and phase transitions between these two structures may occur at high pressures \cite{Haas19651225,Asbrink1999}. According to \citeasnoun{Hautier2010}, ZnNi$_2$O$_4$ is predicted to have a distorted spinel structure (I4$_1$/amd), which has a slightly off-cubic tetragonal unit cell, while ZnCu$_2$O$_4$ is predicted to have a normal cubic spinel structure (Fd$\bar{3}$m). Both structures and their Brillouin zones are shown in Figures \ref{cubicstructure} and \ref{tetragonalstructure}. \citeasnoun{Zhang2010} developed a diagrammatic separation method of classifying BA$_2$X$_4$ compounds into crystal structure types based on the tabulated pseudopotential radii of A and B; using this method, we confirm that both ZnNi$_2$O$_4$ and ZnCu$_2$O$_4$ adopt spinel crystal structures. As both compounds lie on the boundary between normal and inverse spinel structures (Figure 3 of \citeasnoun{Zhang2010}), we hypothesize that they may exhibit similar properties to ZnCo$_2$O$_4$.

\section{Focus of Study}
In this study, we will calculate properties that determine the figure of merit for a transparent conductor \cite{Haacke1976} in order to facilitate the optimization of these properties for the class of compounds under investigation. The figure of merit is defined as:
\begin{equation}
\phi_{TC}=T^x/R_s
\end{equation}
where $T$ is the (optical) transmission (i.e., inverse of absorbance) of a sheet of thickness $t$, $R$ is its resistance, and $x>0$. In terms of the electrical conductivity $\sigma$, the absorption coefficient $\alpha$, and the film thickness that maximizes $\phi_{TC}$, the equation is transformed into:
\begin{equation}
\phi_{TC}=\sigma t \exp \left(-10\alpha t \right)
\end{equation}
and
\begin{equation}
t_{max}=\frac{1}{10\alpha}
\end{equation}

By calculating $\sigma$ and $\alpha$ as a function of structure and chemical composition we will gain atomistic-level insight into how selecting the A atom may affect the performance of zinc oxide spinels; specifically, we are interested in determining whether the maximum theoretically attainable conductivities are sufficient for transparent conducting applications.  We also predict the magnetic behavior of these compounds to ascertain whether they may be classified as dilute magnetic semiconductors.

\section{Calculations}

The electronic density functions and the optical matrix elements are computed with density functional theory \cite{hohenberg_physrev_1964,kohn_physrev_1965}, using a modified version\footnote{Our modification directly outputs the diagonal elements of the optical matrix.} of the Vienna \emph{ab initio} Simulation Package (VASP) \cite{Kresse_CompMatSci_VASPRef0,Kresse_PRB_VASPRef1,Kresse_CompMatSci_VASPRef2,Kresse_PRB_VASPRef3}. Both the generalized gradient approximation (PBE) \cite{Perdew_PRLet_PBEGGA,Perdew_PRLet_PBEGGAErrata} and the hybrid functional with an optimized screening parameter (HSE06) \cite{heyd:8207,heyd:1187,heyd:219906,Krukau2006} are used to represent the exchange-correlation term in the energy functional. PAW potentials \cite{Bloechl_PRB_PAWMethod,Kresse_PRB_PSSToPAW} are used to maintain a balance between accuracy and computational costs. The blocked Davidson iterative matrix diagonalization algorithm \cite{davidson_methcomputmolphys_1983} is used to optimize the orbitals, and the tetrahedron method with Bl\"{o}chl corrections is used to determine partial occupancies. For the hybrid functional calculations, the range separation parameter is set to 0.2 a.u.

To determine the effect of band gap underprediction on property calculations, we apply the scissor operator to the band energies of both DFT and HSE calculation results. By artificially shifting the valence bands and conduction bands symmetrically around the midgap to match the experimental band gaps, we can quantify the effects of the difference in curvature of valence band edge states.

We use two methods to calculate the group velocities of electrons for the determination of material properties. The first method, implemented in the BoltzTraP package \cite{Madsen_CPC_Boltztrapp}, applies Fourier transform expansion of the band energies to determine the gradient along the energy bands, and hence the group velocities of states. Although this method has been proven to yield accurate results across various systems, it is computationally expensive as it requires high $k$-point meshes for convergence.

The second approach to calculating group velocities is to use the Momentum Matrix (PAW-MM) method \cite{PhysRevB.86.115211} to derive the group velocities of the electrons directly from the wavefunction descriptors, which avoids band crossing errors. The velocities are computed directly by the electronic structure code, and are expressed in terms of the optical matrix elements \cite{Scheidemantel_PRB_TransportCoeffsFirstPrinciples}:
\begin{eqnarray} 
\label{momentum_method}
v_{\alpha\beta} \left( i,k \right) & = & \frac{1}{m} \langle \psi_{\alpha\beta} \left( i,k \right) \left| \hat{p} \right| \psi_{\alpha\beta} \left( i,k \right) \rangle
\end{eqnarray}
Here, $v$ are the velocities, $\psi$ are the wavefunctions, $m$ is the electron mass, and $\hat{p}$ is the momentum operator. $\left( i,k \right)$ denotes the eigenvalue-band index and $\alpha\beta$ denote the directional coordinates. The resulting matrices of transition velocities indexed by $k$-value are called the optical matrices. Although this equation may be used to determine group velocities from exact wavefunction descriptions, the PAW approach requires the addition of an augmentation factor to account for errors arising from the pseudization of the wavefunctions around the atomic cores \cite{Furthmuller_PRB_OpticalPropertiesUsingPAW,Kageshima_PRB_MomMatrixCalcPP,Gajdos_PRB_OpticalPropertiesUsingPAW}. 

Once the group velocities are determined, the bulk semi-classical transport properties can be calculated as a function of temperature $T$ and chemical potential $\mu$, using Boltzmann transport theory \cite{Madsen_CPC_Boltztrapp}:
\begin{eqnarray}
\label{eqs_properties1} \sigma_{\alpha\beta} \left( i,k \right) & = & e^2 \tau_{i,k} v_\alpha \left( i,k \right) v_\beta \left( i,k \right) \\
\label{eqs_properties2} \sigma_{\alpha\beta} \left( \varepsilon \right) & = & \frac{1}{N} \sum_{i,k} \sigma_{\alpha\beta} \left( i,k \right)\frac{\delta \left( \varepsilon-\varepsilon_{i,k} \right)}{d\varepsilon} \\
\label{eqs_properties3} \sigma_{\alpha\beta} \left( T; \mu \right) & = & \frac{1}{\Omega} \int \sigma_{\alpha\beta} \left( \varepsilon \right) \left[ -\frac{\partial f_\mu \left(T;\varepsilon \right)}{\partial\varepsilon} \right] d\varepsilon \\
\label{eqs_properties4} \nu_{\alpha\beta} \left( T; \mu \right) & = & \frac{1}{eT\Omega} \int \sigma_{\alpha\beta} \left( \varepsilon \right) \left( \varepsilon-\mu \right) \left[ -\frac{\partial f_\mu \left( T; \varepsilon \right)}{\partial\varepsilon}\right] d\varepsilon   \\
\label{eqs_properties5} S_{\alpha\beta} \left( T; \mu \right) & = & \frac{\nu_{\alpha\beta} \left( T; \mu \right)}{\sigma_{\alpha\beta} \left( T; \mu \right)} 
\end{eqnarray}
Here, $N$ is the number of $k$-points sampled in the IBZ, and is weighted according to the unique vectors generated by the symmetry operators, $\delta$ is the unit impulse function, $\Omega$ is the volume of the primitive cell, $\varepsilon_F$ is the Fermi level, and $f$ is the Fermi-Dirac distribution for an electron gas:
\begin{eqnarray}\label{fermidirac}
f_0 \left( \varepsilon \right) & = & \left[ \exp \left( \frac{\varepsilon - \varepsilon_{F}}{k_B T} \right) + 1 \right]^{-1}
\end{eqnarray}
Equations \ref{eqs_properties3}-\ref{eqs_properties5} yield second-rank tensors for the electrical conductivity, $\sigma$, and Seebeck coefficient, $S$; the latter is used to determine whether the material is p-type ($S>0$) or n-type ($S<0$). To quantify these properties as scalars, the trace of each matrix is evaluated.  While all calculations of electronic wave functions are performed for the ground state (e.g., 0 K), electronic properties at higher temperatures are simulated by applying the Fermi distribution over electronic states, as described in Equation \ref{fermidirac}.

By observing the area around the Fermi level for any computed structure, we can obtain a picture of the effect of light doping on that structure. We assume that such doping can be achieved without significant distortion of the native band structure; this is the rigid band approximation (RBA).

After extensive $k$-point sampling convergence tests for the BoltzTraP method, the BZ for the Fd$\bar{3}$m system is sampled with a $41\times41\times41$ $k$-point grid, which yields 1771 unique $k$-points in the IBZ, while the BZ for the I4$_1$amd system is sampled with a $28\times28\times28$ $k$-point grid, which yields 1639 unique $k$-points in the IBZ. In order to save computational costs, we perform all calculations simultaneously with this sampling rate. The kinetic energy cutoff for the plane wave basis set is set to 520 eV. The resulting wavefunctions are used in the Momentum Matrix equation, with a PAW potential correction to determine the group velocities. Equations \ref{eqs_properties1}-\ref{eqs_properties5} are then used to determine the transport properties from the group velocities with an in-house MATLAB \cite{matlab} code.

\section{Structure Optimization}

\begin{table}
\begin{minipage}{\textwidth}
\caption{\label{energy_table}Total energy per primitive cell (14 atoms) of all systems studied. Lattice parameters are for the corresponding unit cell.}
\footnotesize\rm
\begin{tabular*}{\textwidth}{@{}l*{15}{@{\extracolsep{0pt plus12pt}}l}}
\br
System&Space group&Magnetism&Energy(eV)&lattice parameter(s) ({\AA})\\
\hline\hline
ZnCu$_2$O$_4$&Fd$\bar{3}$m&NFM&-93.816&\\
\space	&&AFM&-98.023&\\
\space	&&FM&-98.369&\textit{a}=8.291\\
\hline
\space	&I4$_{1}$amd&NFM&-95.598\\
\space	&&AFM&-98.024\\
\space	&&FM&-98.369&\textit{a}=5.862,\textit{c}=8.296\\
\hline\hline
ZnNi$_2$O$_4$&Fd$\bar{3}$m&NFM&--\\
\space	&&AFM&-100.666\\
\space	&&FM&-100.343\\
\hline
\space	&I4$_{1}$amd&NFM&-112.721\\
\space	&&AFM&-116.693&\textit{a}=7.912,\textit{b}=7.922,\textit{c}=8.836\\
\space	&&FM&-116.399&\textit{a}=5.942,\textit{c}=7.797\\
\hline\hline
ZnCo$_2$O$_4$&Fd$\bar{3}$m&NFM&-130.863&\textit{a}=8.081\\
\space	&&AFM&-130.864&\textit{a}=8.077\\
\space	&&FM&-127.883\\
\br
\end{tabular*}
\end{minipage}
\end{table}

The results of the HSE06 structural optimizations are shown in Table \ref{energy_table}. Both the cubic and tetragonal structures were evaluated for the two new compounds, ZnCu$_2$O$_4$ and ZnNi$_2$O$_4$, in three possible spin configurations (i.e., nonmagnetic (NFM), antiferromagnetic (AFM), and ferromagnetic (FM)). However, for the previously characterized compound, ZnCo$_2$O$_4$, only the known cubic structure was investigated. Only the lattice parameters of the two most stable structures for each compound are showns. ZnCo$_2$O$_4$ exhibits two possible structures, NFM and AFM, which suggests that this compound is not likely to possess desirable properties as a DMS. Comparison to the experimental lattice parameter of 8.0946(2) {\AA} (JCPDS card no. 23-1390) \cite{sharma2007nanophase} suggests that the NFM configuration is more representative of the experimentally observed structure. ZnNi$_2$O$_4$ is most stable as an AFM structure, with its FM structure being 0.3 eV less stable. ZnCu$_2$O$_4$ shows the most promise as a DMS, with the cubic and the tetragonal FM structures being equally stable. The identical total energy of these two structures suggests that fabrication method and substrate are important in determining the preferred structure formed for this compound.

Table \ref{energy_table_inversion} shows the total energy for the cubic ZnCu$_2$O$_4$ primitive cells under partial and complete inversion, and the corresponding lattice parameters of the unit cell for the most stable systems. As shown, the FM system is always the most thermodynamically configuration, and the inverted spinel structure is less stable than the normal spinel.

\begin{table}
\begin{minipage}{\textwidth}
\caption{\label{energy_table_inversion}Total energy per primitive cell (14 atoms) of cubic ZnCu$_2$O$_4$ systems under partial and total inversion. Lattice parameters are for the corresponding unit cell.}
\footnotesize\rm
\begin{tabular*}{\textwidth}{@{}l*{15}{@{\extracolsep{0pt plus12pt}}l}}
\br
System&Symmetry&Magnetism&Energy(eV)&lattice parameter(s) ({\AA})\\
\hline\hline
50\% inversion&R3m&NFM&-94.722&\\
\space	&&AFM&-96.529&\\
\space	&&FM&-96.807&\textit{a}=5.843,\textit{c}=14.659\\
\hline
100\% inversion&Imma&NFM&--\\
\space	&&AFM&-95.286\\
\space &&FM&-95.577&\textit{a}=5.826,\textit{a}=6.103,\textit{c}=8.140\\
\br
\end{tabular*}
\end{minipage}
\end{table}

\begin{table}
\begin{minipage}{\textwidth}
\caption{\label{O_B_distance_table}Mean distance between octahedrally coordinated atoms, Zn and B, and the corresponding O atoms.}
\footnotesize\rm
\begin{tabular*}{\textwidth}{@{}l*{15}{@{\extracolsep{0pt plus12pt}}l}}
\br
System&Symmetry&Magnetism&Mean B-O distance ({\AA})&Mean Zn-O distance ({\AA})\\
\hline\hline
ZnCo$_2$O$_4$&Fd$\bar{3}$m&AFM&2.24&\\
ZnCo$_2$O$_4$&Fd$\bar{3}$m&NFM&2.24&\\
\hline\hline
ZnCu$_2$O$_4$&Fd$\bar{3}$m&FM&2.34&\\
ZnCu$_2$O$_4$&I4$_1$amd&FM&2.42&\\
\hline\hline
ZnCu$_2$O$_4$ 50\% inversion&R3m&FM&2.35&2.52\\
ZnCu$_2$O$_4$ 100\% inversion&Imma&FM&2.38&2.50\\
\hline\hline
ZnNi$_2$O$_4$&I4$_1$amd&FM&2.33&\\
ZnNi$_2$O$_4$&Fddd&AFM&2.33&\\
\br
\end{tabular*}
\end{minipage}
\end{table}

Table \ref{O_B_distance_table} shows the mean distance between the Zn and B octahedrally coordinated atoms and the corresponding O atoms. As the atomic number of B increases, so does the mean distance, independent of the structure. Inversion in cubic ZnCu$_2$O$_4$ system leads to increased distance between the Cu atom and its O neighbors. Also, from these systems we see that, as expected, Zn-O distances are greater than for any of the B atoms, and that increased inversion leads to O atoms being more tightly packed around the Zn atom. From the Co and Ni systems, we see that the most stable structures have the same mean B-O distances, regardless of magnetism. From the Cu system, we see that the tetragonal structure has greater B-O distances than the cubic structure. This will likely result in weaker hybridization of orbitals in the tetragonal structure than in the cubic structure.

\section{Optical Properties}
\begin{figure}
\includegraphics[width=475pt]{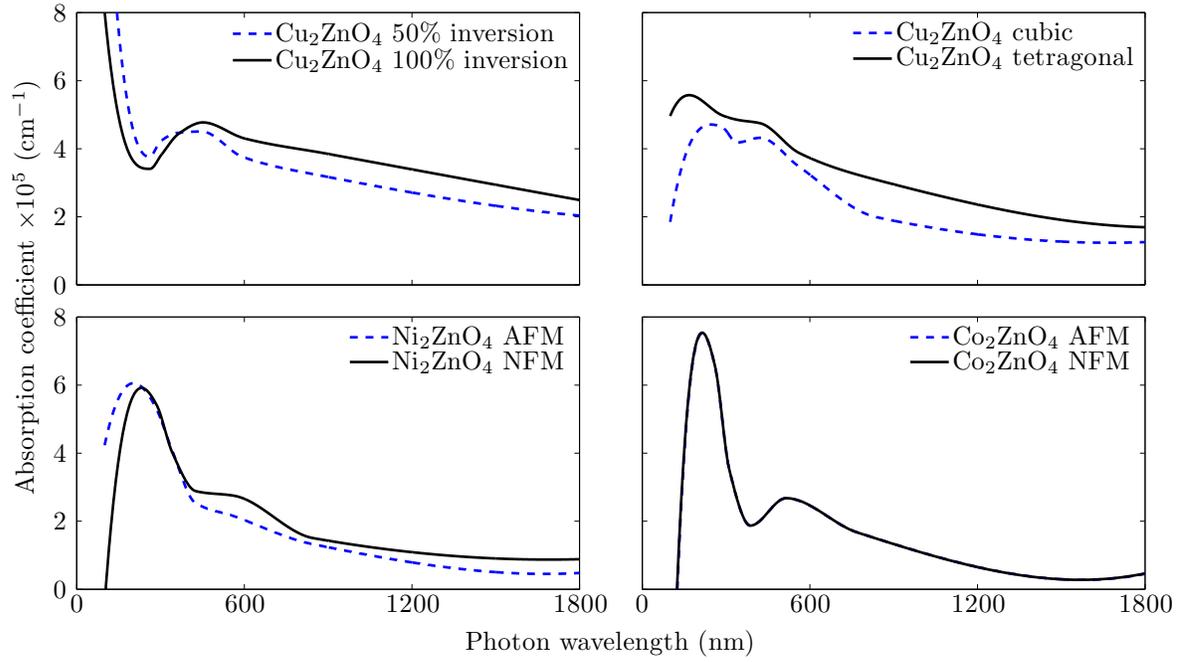}
\caption{Absorption coefficient for all systems studied.}\label{all_absorption}
\end{figure}

The absorption coefficient spectra for the bulk materials are shown in Figure \ref{all_absorption}. The absorption in the visible range (390-700 nm) is high for all three compounds. ZnNi$_2$O$_4$ exhibits an absorption coefficient similar to that of ZnCo$_2$O$_4$, although it is higher in the ultraviolet range. Interestingly, the slightly less stable AFM structure has a lower absorption throughout the visible range, suggesting that if this structure could be preferentially synthesized, a more transparent compound would be formed. ZnCu$_2$O$_4$ has a much higher absorption throughout the calculated spectrum. The cubic structure has considerably lower absorption for $\lambda > 600 \ \textrm{nm}$, so again, if the synthesis could be controlled in favor of cubic lattice formation, a much more transparent compound would be formed. Cation inversion in the cubic ZnCu$_2$O$_4$ structure seems to decrease absorption at lower wavelengths, which may pose an option for engineering more transparent spinels.  However, as the percentage of inversion increases, absorption increases as well for $\lambda > 600 \ \textrm{nm}$.  Therefore, should this approach be used to decrease absorption for low $\lambda$, we should consider all of these effects to account for increased absorption at higher wavelengths.  It may still be possible to create very thin films of these materials, so that their optical absorption is not an impediment to transparent conducting applications. 

\section{Electronic Properties}

\subsection{Band Structures}

\begin{figure}
\includegraphics[width=230pt,angle=270]{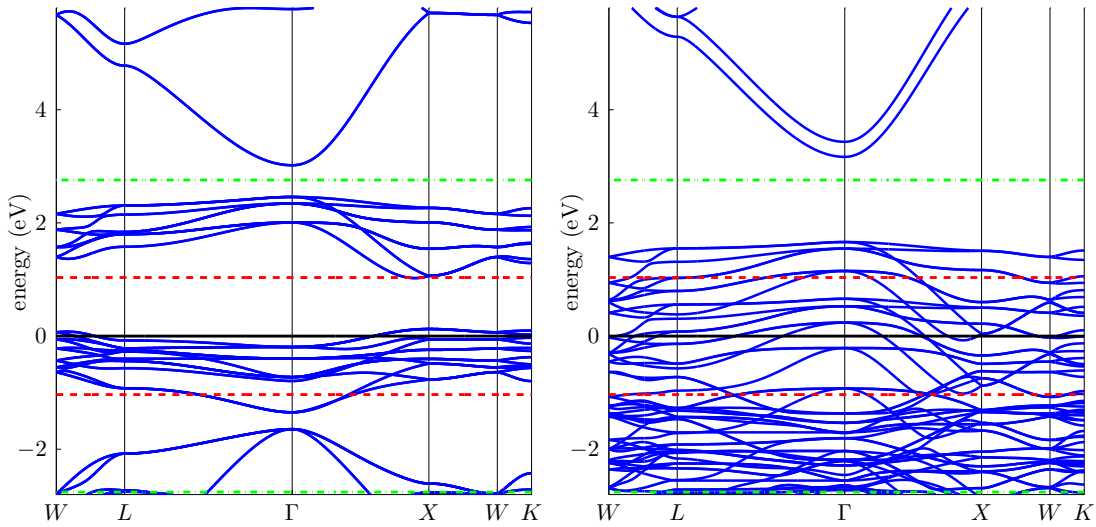}
\caption{DFT band structures for ZnCo$_2$O$_4$ (left), for which the band gap is underestimated, and ZnCu$_2$O$_4$ (right), for which the conduction and valence bands are entangled. \label{comparison_dft_cocu}}
\end{figure}

For all systems (except ZnCo$_2$O$_4$) computed with DFT, the electronic band gap entirely disappears, so that the conduction and valence bands become entangled (see Figure \ref{comparison_dft_cocu}). This is due to the well-known band gap underestimation caused by the incorrect treatment of electron correlation-exchange in DFT. As a result, the Fermi level falls within one long, extended band, and the material appears to be metallic. \footnote{The red and green dashed lines in all band structure plots show the filter of the distribution function, (Equation \ref{fermidirac}), at 300 K and 800 K, respectively; these are the bands around the Fermi level that contribute to the summation in the conductivity calculation. Because of the exponential term in this expression, band energy contributions outside this distribution are zero. The filter for 300 K is much smaller than for 800 K. This result has two consequences: 1. For low temperatures, fewer bands are included in the summation, and thus a higher $k$-point mesh is necessary to obtain accurate estimations of the electrical properties, and 2. High temperature calculations are much more sensitive to errors in the determination of the band gap.} Therefore, DFT is inadequate for predicting the electronic properties of the ternary oxides under study, and is not suitable for screening these materials. Later, we will quantify whether recalculating the band gap at the $\Gamma$ point with a method that more accurately describes the exchange correlation interactions, such as hybrid DFT with the HSE06 functional, and then applying the scissor operator is sufficient to correctly describe the band curvature and thus determine electronic properties.

\begin{figure}
\includegraphics[width=420pt]{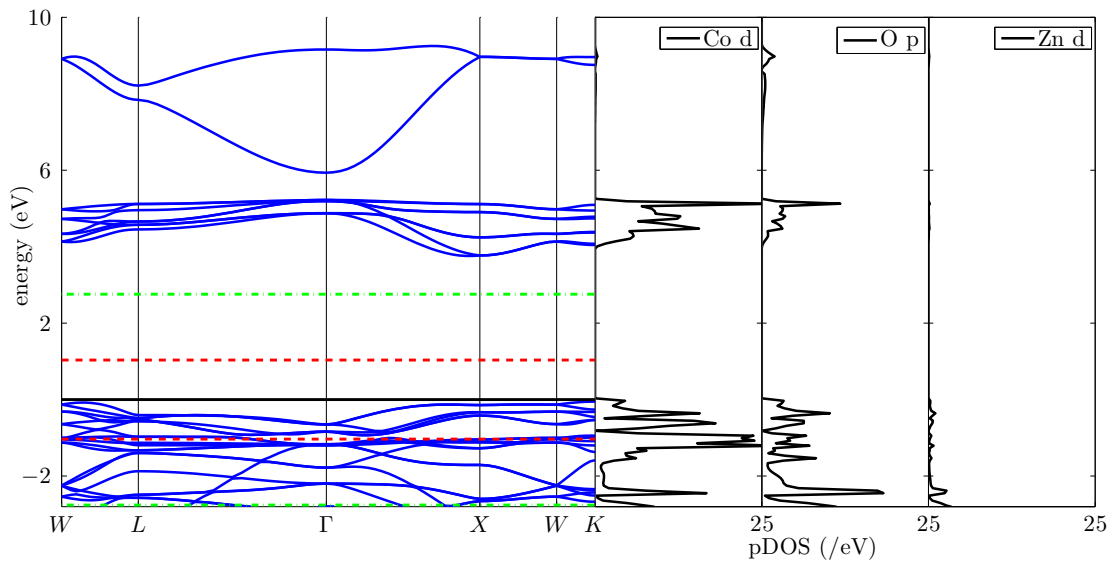}
\caption{Band structure and partial density of states (pDOS) for cubic NFM ZnCo$_2$O$_4$.\label{co2zno4hse_NFM_bsdos}}
\end{figure}
\begin{figure}
\includegraphics[width=420pt]{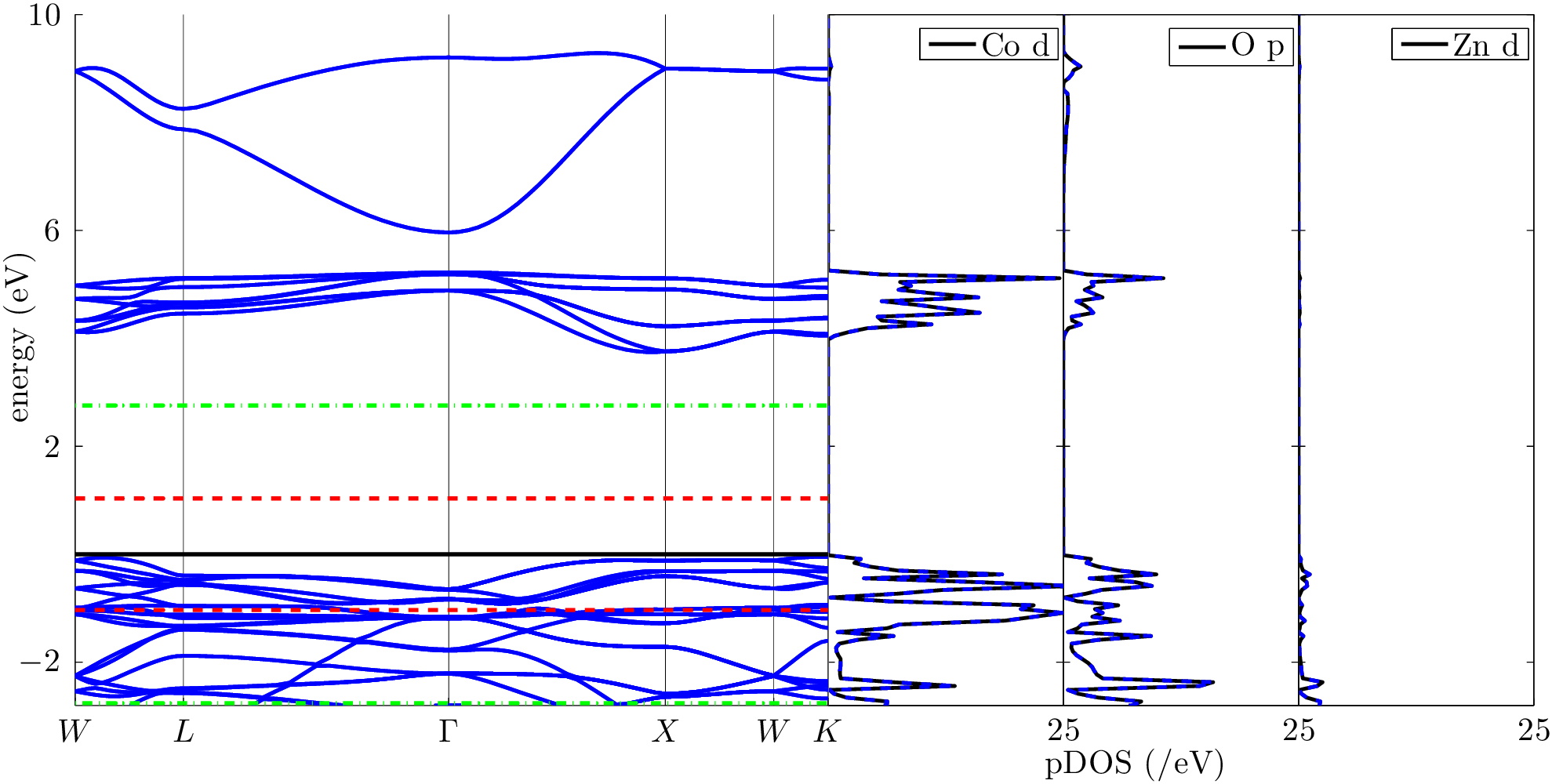}
\caption{Band structure and partial density of states (pDOS) for cubic AFM ZnCo$_2$O$_4$.\label{co2zno4_bsdos}}
\end{figure}

\begin{figure}
\includegraphics[width=420pt]{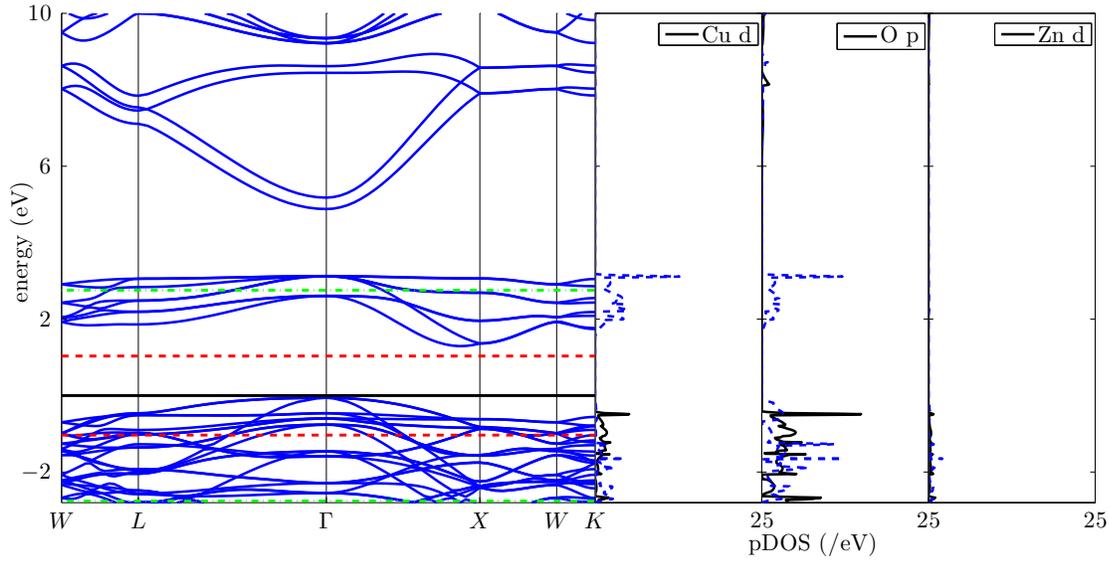}
\caption{Band structure and partial density of states (pDOS) for cubic FM ZnCu$_2$O$_4$.\label{cu2zno4_bsdos}}
\end{figure}

\begin{figure}
\includegraphics[width=420pt]{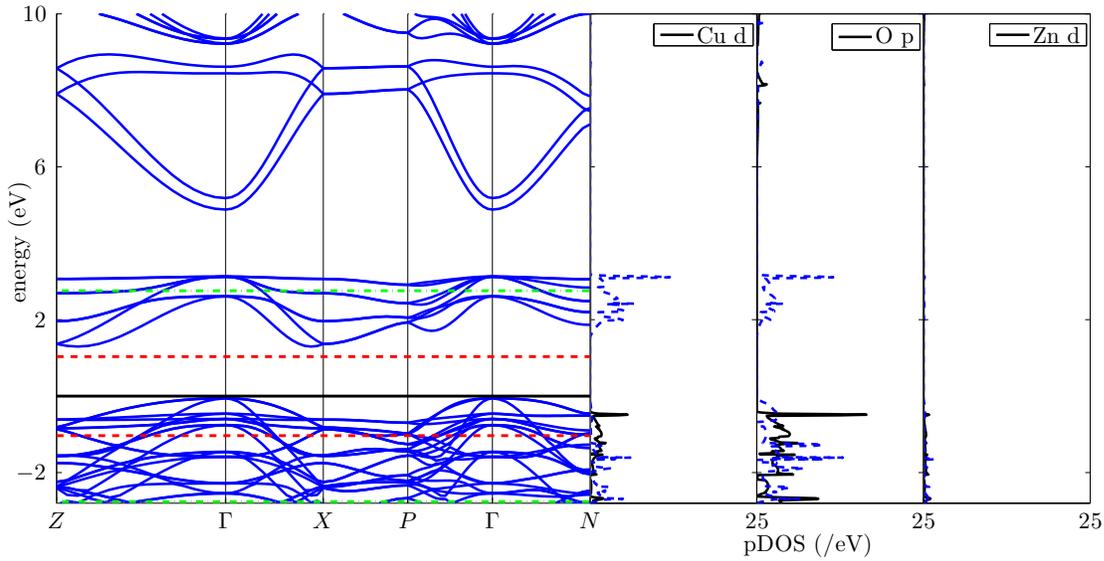}
\caption{Band structure and partial density of states (pDOS) for tetragonal FM ZnCu$_2$O$_4$.\label{cu2zno4_tFM_bsdos}}
\end{figure}

\begin{figure}
\includegraphics[width=420pt]{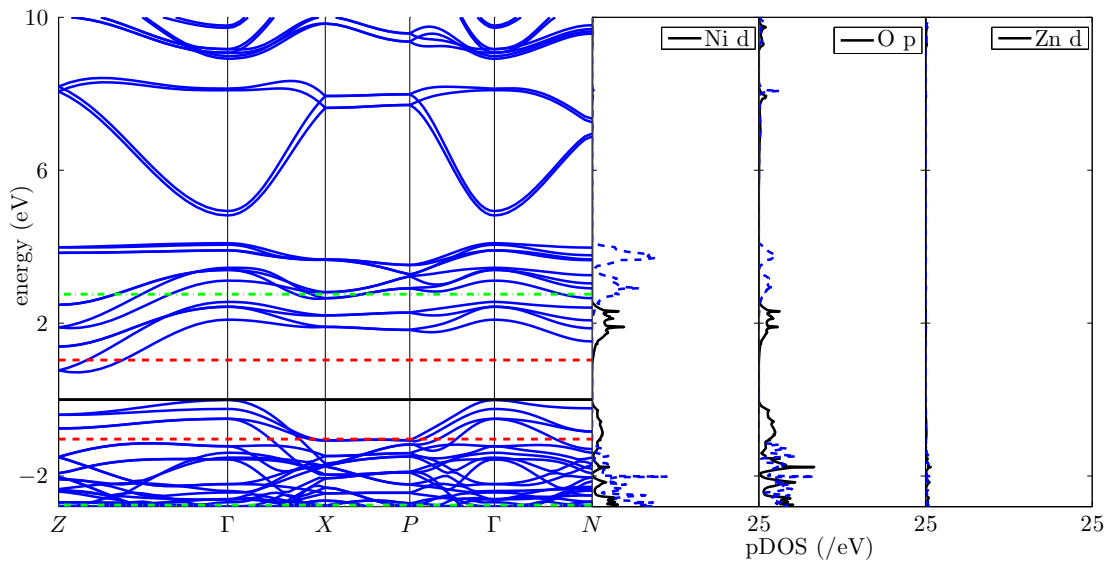}
\caption{Band structure and partial density of states (pDOS) for distorted tetragonal spinel (I4$_{1}$amd) FM ZnNi$_2$O$_4$.\label{ni2zno4_bsdos}}
\end{figure}

\begin{figure}
\includegraphics[width=420pt]{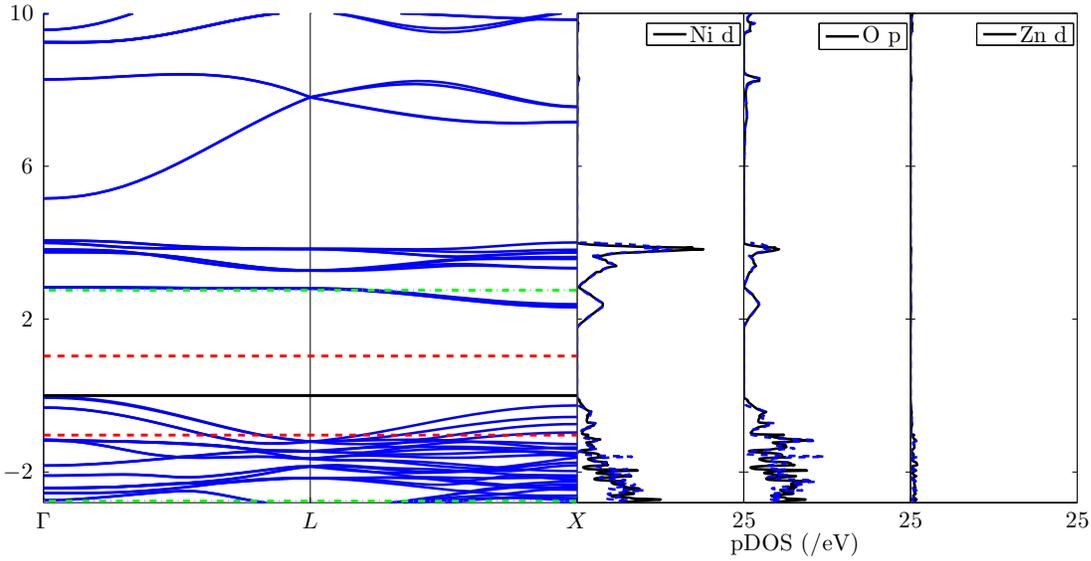}
\caption{Band structure and partial density of states (pDOS) for distorted orthorhombic spinel (Fddd) AFM ZnNi$_2$O$_4$.\label{ni2zno4_fddd_bsdos}}
\end{figure}

Figures \ref{co2zno4_bsdos}-\ref{ni2zno4_fddd_bsdos} show the HSE06 band structures and partial density of states (pDOS) for the systems under consideration with the lowest total energies. In all spin polarized pDOS calculations, the solid line represents spin up states, and the dotted line represents spin down states. For ZnCo$_2$O$_4$, the indirect band gap is 3.72 eV, which represents an increase of 2.82 eV from the DFT calculation, and an increase of 0.9 eV from the value reported by \citeasnoun{Paudel2011doping}. From the pDOS, we see that the O 2p orbitals contribute considerably to the valence band, with a significant shallow hybridization with Co 3d orbitals and a deep valence hybridization with the Zn 3d orbitals (not shown). Spin up and spin down state populations in the AFM structure are perfectly aligned.

For the other two systems, ZnCu$_2$O$_4$ and ZnNi$_2$O$_4$, an interesting phenomenon occurs; namely, the density of states shows dependency on the spin polarization. The conduction band, which is similar in shape to the one observed in ZnCo$_2$O$_4$, is occupied by only spin down electrons in ZnCu$_2$O$_4$. In tetragonal ZnNi$_2$O$_4$, the conduction band is split between spin up and spin down states. This band is located much closer to the valence band; the calculated band gaps for the Cu and Ni compounds are significantly smaller, being 1.17 eV for ZnCu$_2$O$_4$ and 1.67 eV for ZnNi$_2$O$_4$. The favorable band gap and spin polarization distribution results in ZnCu$_2$O$_4$ and tetragonal FM ZnNi$_2$O$_4$ being ideal for magnetoelectronics applications. 

Both the cubic and tetragonal FM ZnCu$_2$O$_4$ systems have a valence band that is populated by both spin up and spin down states, mainly of O 2p character, with some 3d hybridization from the Cu valence. This material will exhibit polarized conductivity and a band gap of 1.17 eV. Under a magnetic field, this material can be switched to exhibit properties between those of an insulator and semiconductor. The insulating band gap for spin up electrons is 5.2 eV. The valence band states have significant curvature as compared to the ZnCo$_2$O$_4$ compound at the $\Gamma$ point, which signals low electron effective mass and potentially high p-type semiconducting behavior.

The tetragonal FM ZnNi$_2$O$_4$ system has only spin up electrons occupying shallow states of the top of the valence band and bottom of the conduction band. The spin up effective band gap is 1 eV and the spin down effective band gap is 3 eV, which makes this material attractive for applications where the material band gap needs to be dynamically tuned. The distorted structure stabilizes the spin down electron states in the conduction band, bringing them closer to the valence band. Whereas the valence band edge in the cubic structure is relatively flat (not shown), it shows higher curvature in the distorted spinel, which indicates high carrier mobility. All of these factors may explain why the distorted structure is thermodynamically much more favorable in ZnNi$_2$O$_4$. The orthorhombic AFM spinel ZnNi$_2$O$_4$ structure shows a wider indirect band gap (2.2 eV) than the other new alloys, strong curvature of the valence band, and a relatively flat conduction band edge. All of these factors make this material ideal for transparent conducting applications.

It is evident that in all of these materials, the octahedrally coordinated Zn atom is relatively inert in determining the valence electronic properties, and it is the interaction between the tetrahedrally coordinated B 3d electrons and the O 2p electrons that strongly influences these properties in the compound. With increasing atomic number of B, we notice a significant narrowing of the band gap. Additionally, we note that with increasing atomic number of B, there is less contribution from the B 3d orbitals to the shallow valence band states, resulting in less stabilization of the O 2p orbitals, and a broader hybridization range throughout the shallow valence states. We also notice that the curvature of the valence band also increases from ZnCo$_2$O$_4$ to ZnNi$_2$O$_4$ and ZnCu$_2$O$_4$, mainly due to the destabilized O 2p states; this may possibly lead to higher electrical conductivities upon light p-type doping in the latter two compounds. Therefore, there is a correlation between the destabilization of O 2p orbitals and greater curvature of the valence band edge, wherein the properties of the material can be optimized with the choice of element B. 

\subsection{Cation Inversion}

\begin{figure}
\includegraphics[width=420pt]{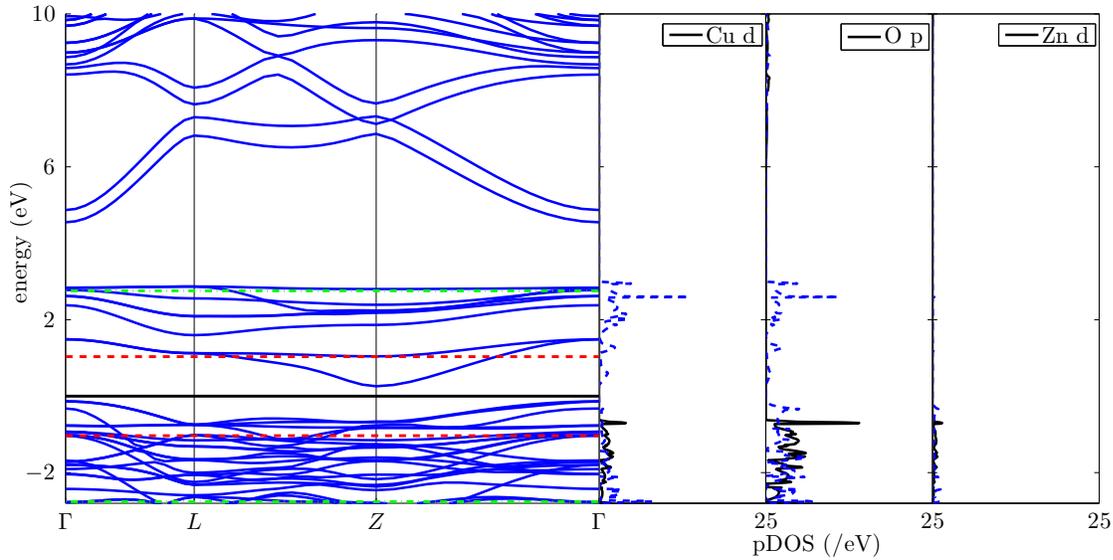}
\caption{Band structure and partial density of states (pDOS) for cubic spinel FM ZnCu$_2$O$_4$ with 50 \% cation inversion.\label{cu2zno4_inv50_bsdos}}
\end{figure}
\begin{figure}
\includegraphics[width=420pt]{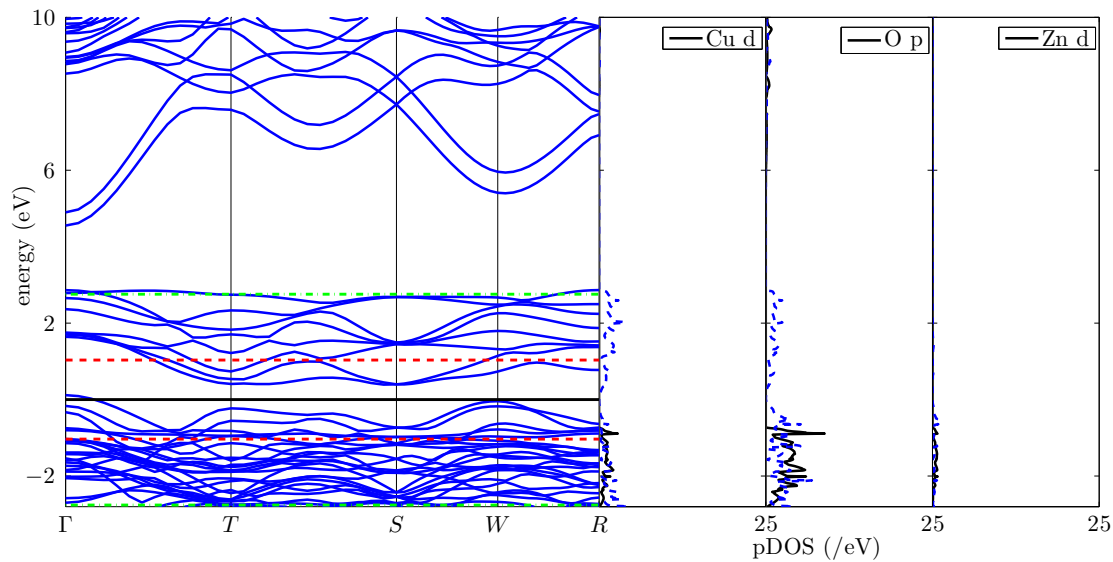}
\caption{Band structure and partial density of states (pDOS) for cubic spinel FM ZnCu$_2$O$_4$ with 100\% cation inversion.\label{cu2zno4_inv100_bsdos}}
\end{figure}

When synthesizing spinels, one of the most common distortions in the structure is cation inversion, or the preferential swapping of B atoms for Zn atoms. Since we have determined that the 3d electrons of the octahedrally coordinated B atom strongly influence electronic properties, it is important to study cation inversion to determine its effect on the band structure of the material. Although small fractions (i.e., $< 10 \textrm{\%}$) of inversion are achievable in practice, it is too computationally expensive to perform calculations for these small fractions of inversion; however, we note that the rigid band approximation we use for doping can also be used to approximate very small fractions of inversion, and can be treated as a form of light self-doping within the material. By looking at the trends from 0 to 50 to 100\% inversion, we can determine the electronic characteristics of inversion, and thus, the impact on the properties of the materials. 

We look at cubic ZnCu$_2$O$_4$ as an illustrative example of the effect of cation inversion on material properties. We find that, again, for 50\% and 100\% cation inversion, the FM structure is the most stable, so we determine that this compound is likely ferromagnetic regardless of atom arrangement. Additionally, by examining the band structures (see Figures \ref{cu2zno4_inv50_bsdos}-\ref{cu2zno4_inv100_bsdos}), we notice that inversion causes narrowing of the band gap, and thus, a likely increase in electrical conductivity but a decrease in transparency. As indicated by the increased average distances of the octahedrally coordinated Cu and Zn atoms, the hybridization peak of the transition metal 3d electrons and the O 2p electrons at the top edge of the valence band is much smaller than in the normal structure. Furthermore, inversion may possibly hinder p-type doping capabilities.  Again, although such high levels of cation inversion are unlikely in practice, as corroborated by the total energy data, these computations lead us to conclude that cation inversion has adverse effects on transparency, and the occurrence of such inversion should be minimized during fabrication. 

\subsection{Bader Charge Analysis}

Table \ref{bader_analysis} shows the calculated electronic charges of B and Zn atoms on their native sites in the normal spinel structure, and at a defect site, where stoichiometry of the lattice is preserved (i.e., cation inversion). The energy levels of these defects introduce donor and acceptor levels in the material, and determine whether the Fermi level is pinned \cite{Paudel2011doping}. We are interested in qualitatively describing what happens to the electronic structure as defects are introduced into the lattice. As indicated in Table \ref{bader_analysis}, as the atomic number of B increases, the charge at the Zn-site also increases slightly. Thus, the added charge is dislocated towards the Zn atom, which stabilizes the charge at a deep valence state. According to this analysis, when the B and Zn atoms swap positions and the stoichiometry of the system is maintained, the atoms also swap formal charges. This indicates that donor and acceptor levels do not compensate each other, so doping is permitted.

\begin{table}
\begin{minipage}{\textwidth}
\caption{\label{bader_analysis}Electronic charges in systems with maintained stoichiometry.}
\footnotesize\rm
\begin{tabular*}{\textwidth}{@{}l*{15}{@{\extracolsep{0pt plus12pt}}l}}
\br
Atom&Atomic Number(Valence)&Native Location&Native Charge&Defect Charge\\
\mr
ZnCo$_2$O$_4$\\
\space	Co&27(9)&T$_d$&7.56&7.56/10.52\footnote{non-native site charge}\\
\space	Zn&30(12)&O$_h$&10.68&10.69/7.72\footnotemark[\value{footnote}]\\
\space	O&8(6)& &7.05&7.05\\
\hline
ZnNi$_2$O$_4$ (I4$_1$amd)\\
\space	Ni&28(10)&T$_d$&8.40&8.58/10.58\footnotemark[\value{footnote}]\\
\space	Zn&30(12)&O$_h$&10.69&10.70\footnotemark[\value{footnote}]/8.67\\
\space	O&8(6)& &7.12&7.04\\
\hline
ZnCu$_2$O$_4$ \\
\space	Cu&29(11)&T$_d$&9.58&9.59/10.56\footnotemark[\value{footnote}]\\
\space	Zn&30(12)&O$_h$&10.73&10.73/9.80\footnotemark[\value{footnote}]\\
\space	O&8(6)& &7.03&7.02\\
\br
\end{tabular*}
\end{minipage}
\end{table}

Table \ref{bader_analysis2} shows a second analysis that was performed on the charge around the B and Zn atoms in the structure. In this second analysis, the stoichiometry of the structure is changed (i.e., self-doping occurs); in the first calculation one B atom was used to replace Zn and in the second calculation one Zn atom was used to replace B. As shown in the tabulated values, in these non-stoichiometric systems, the formal charge on the atoms does not change significantly when their position in the lattice changes, although there is a small change. A atoms gain some charge when located at O$_h$ sites, while Zn loses net charge at the T$_d$ site. This latter effect indicates that the stabilizing effect of Zn is negated, and thus free carrier density increases. This effect is desirable and useful when engineering a self-doped system.

\begin{table}
\begin{minipage}{\textwidth}
\caption{\label{bader_analysis2}Electronic charges with non-stoichiometric defects.}
\footnotesize\rm
\begin{tabular*}{\textwidth}{@{}l*{15}{@{\extracolsep{0pt plus12pt}}l}}
\br
Atom&Atomic Number(Valence)&Normal charge&B on O$_h$&Zn on T$_d$\\
\mr
ZnCo$_2$O$_4$\\
\space	Co&27(9)&7.56&7.56/7.73\footnote{non-native site charge}&7.56\\
\space	Zn&30(12)&10.68&10.69&10.68/10.51\footnotemark[\value{footnote}]\\
\space	O&8(6)&7.05&7.04&7.05\\
\hline
ZnNi$_2$O$_4$ (I4$_1$amd)\\
\space	Ni&28(10)&8.40&8.61/8.61\footnotemark[\value{footnote}]&8.58\\
\space	Zn&30(12)&10.69&10.71&10.69/10.58\footnotemark[\value{footnote}]\\
\space	O&8(6)&7.12&7.02&7.04\\
\hline
ZnCu$_2$O$_4$ \\
\space	Cu&29(11)&9.58&9.59/9.86\footnotemark[\value{footnote}]&9.57\\
\space	Zn&30(12)&10.73&10.72&10.73/10.53\footnotemark[\value{footnote}]\\
\space	O&8(6)&7.03&7.00&7.03\\
\br
\end{tabular*}
\end{minipage}
\end{table}

\subsection{Electrical Conductivity\label{conductivity}}

In this section we present the conductivity values of the systems studied. The power factor, $S^2\sigma$, is shown to emphasize the fluctuations in the electrical conductivity. Calculated properties are presented for 300 K, and the presented values are those obtained using the Momentum Matrix method. The BoltzTraP method yields similar curves, with a discrepancy in magnitude of less than 5\%. 

\begin{figure}
\includegraphics[width=475pt]{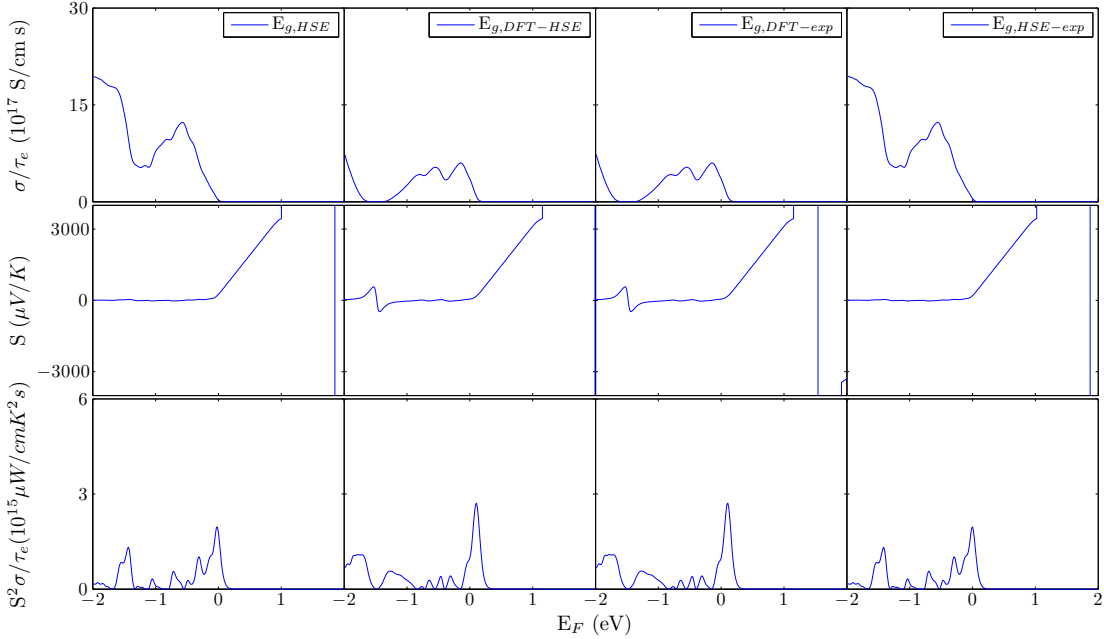}
\caption{Comparison of different methods for calculating the electronic transport properties for the case of ZnCo$_2$O$_4$, for which the DFT valence and conduction bands were not entangled.}\label{compare_co_gap}
\end{figure}

First, we analyze the difference in calculated properties between the DFT and HSE computational approaches. When the band gap is merely underpredicted with DFT (e.g., the ZnCo$_2$O$_4$ system), we rectify this underprediction by applying the scissor operator in the middle of the band gap to obtain the same band gap as that determined by the HSE functional. Furthermore, we also compute $\sigma$ for both the HSE and the DFT results with the reported band gap of 2.8 eV \cite{Paudel2011doping} to determine whether the underestimation or overestimation of this gap affects our results. $\sigma$, $S$, and the power factor, $S^2\sigma$, are shown in Figure \ref{compare_co_gap} for the four scenarios of adjusted band gap: HSE unadjusted (HSE), DFT adjusted to match HSE (DFT-HSE), DFT adjusted to match experiment (DFT-exp), and HSE adjusted to match experiment (HSE-exp). From these results we determine that: 1. Artificially changing the band gap does not discernibly affect the magnitude and distribution of the $\sigma$ values at 300 K, 2. The curvature of the bands from the DFT calculation describes the valence states much differently from the those obtained using the HSE calculation, due to the incorrect exchange interaction factor, and 3. When the HSE calculation is corrected by manually narrowing the band gap to fit previously reported values, the conductivity values are not affected by the band gap difference. These results indicate that the distribution of states in the summation for property calculation is not dependent on the width of the band gap, as long as the conduction bands lie outside the Fermi distribution (this observation is corroborated by the previously presented band structures). The last result also means that we may calculate the electronic properties of these p-type alloys at 300 K using the HSE determined band gap, without adjustment for experiment, as long as our band gap is not over or underpredicted by an amount that would place the bottom conduction band within the Fermi distribution range filter at 300K. Since the computed band gap deviates from the experimentally observed gap by 0.9 eV, we originally expected that the computed band curvature will not be exact. However, the HSE band curvatures are more accurate than the DFT values, as the DFT band gap is underpredicted by a much larger amount (1.9 eV), and underprediction leads to greater deviation of band curvature due to exchange interaction. Therefore, we choose the HSE method to characterize the new systems in the following analysis.

\begin{figure}
\includegraphics[width=475pt]{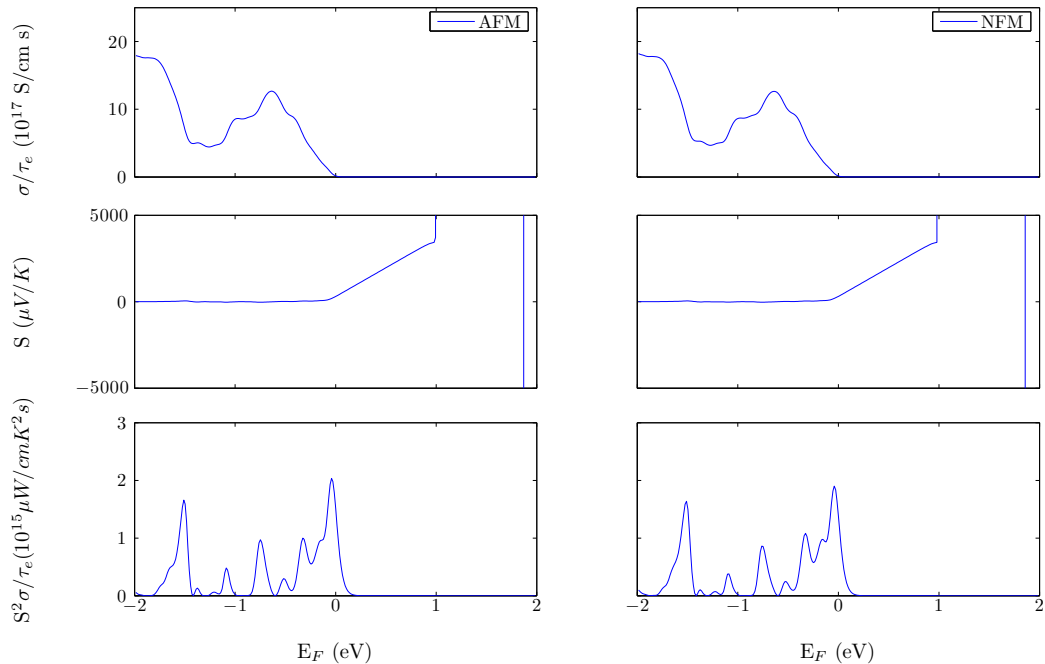}
\caption{Electronic transport properties for ZnCo$_2$O$_4$.}\label{co2zno4_properties}
\end{figure}

Figure \ref{co2zno4_properties} shows the $\sigma$, $S$, and $S^2 \sigma$ values for AFM and NFM cubic ZnCo$_2$O$_4$.\footnote{Here, $S$ approaches infinity from the left and negative infinity from the right, since the material has a wide band gap and is intrinsically an insulator.} We note the first optimum doping level at -0.6 eV for p-type conductivity in this material. Assuming $\tau_e \approx 10^{-14} \ \textrm{s}$\footnote{We make this assumption for all future calculations of $\sigma$ mentioned in the text.} yields a local maximum conductivity of 1.3 $\times$ 10$^{4}$ S/cm with a broad average optimum conductivity of $\sigma = 1 \times 10^{4} \ \textrm{S}/\textrm{cm}$. At 673 K, the optimum conductivity is $\sigma = 1.2 \times 10^{4} \ \textrm{S}/\textrm{cm}$. The intrinsic conductivity at 300 K and 673 K is $\sigma = 1.5 \times 10^{2} \ \textrm{S}/\textrm{cm}$ and $\sigma = 5.5 \times 10^{2} \ \textrm{S}/\textrm{cm}$ for the NFM structure and $\sigma = 1.6 \times 10^{2} \ \textrm{S}/\textrm{cm}$ and $\sigma = 5.8 \times 10^{2} \ \textrm{S}/\textrm{cm}$ for the AFM structure.; this is comparable to the $\sigma = 100 \ \textrm{S}/\textrm{cm}$ reported by \citeasnoun{Paudel2011doping} in their off-stoichiometric "doped'" NiCo$_2$O$_4$. The likely computational sources for this deviation are twofold: 1. The highly off-stoichiometric character of the experimentally deposited sample, and 2. The high temperature at measurement. Both factors would increase the concentration of defects in the material and distort the electronic structure from the theoretical values. We note that the computed Fermi level lies high enough in the band gap that $\sigma$ is intrinsically far removed from its optimum value. A material with a lower lying intrinsic Fermi level would be ultimately preferable.

\begin{figure}
\includegraphics[width=475pt]{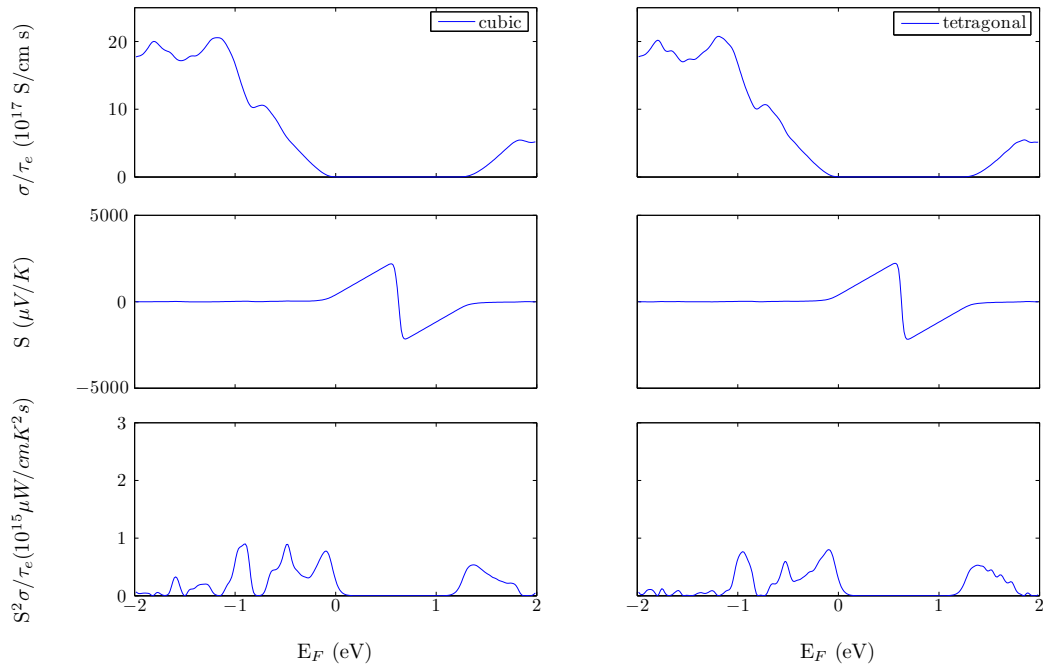}
\caption{Electronic transport properties for ZnCu$_2$O$_4$.}\label{cu2zno4_properties}
\end{figure}

\begin{figure}
\includegraphics[width=475pt]{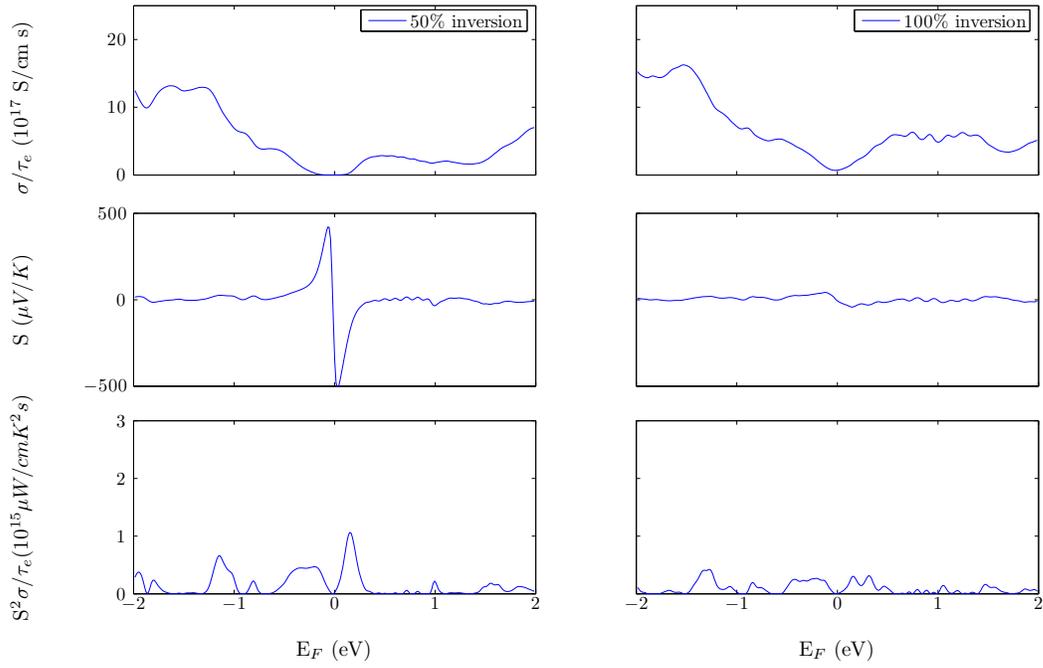}
\caption{Electronic transport properties for cubic ZnCu$_2$O$_4$ with cation inversion.}\label{cu2zno4inv_properties}
\end{figure}

Figure \ref{cu2zno4_properties} shows the electrical properties of cubic and tetragonal ZnCu$_2$O$_4$. Here, the cubic material is not a wide gap insulator, and hence the $S$ value does not saturate within the gap. The undoped conductivity of the material at 300 K is predicted to be $\sigma = 17 \ \textrm{S}/\textrm{cm}$ and the first local minimum is of the same magnitude and occurs at approximately the same doping level (i.e., -0.73 eV for Cu vs. -0.65 eV for Co) as the ZnCo$_2$O$_4$ compound. However, the second local maximum in ZnCu$_2$O$_4$, at $\varepsilon_F = -1.1 \textrm{eV}$, is $\sigma = 2 \times 10^{4} \ \textrm{S}/\textrm{cm}$, which is twice the value of the second local maximum in ZnCo$_2$O$_4$; this peak is much broader and occurs at a lower doping level than in ZnCo$_2$O$_4$. Thus, between -0.9-0 eV, the doping patterns are similar between the two compounds, but for higher doping levels, the Cu-containing compound appears to be twice as conductive as the Co-containing compound.  This result confirms the prediction for higher carrier mobility due to the increased curvature of the bands. As supported by the electronic structure calculation, wide ranges of strong hybridization in shallow valence band states leads to a broader stable conductivity curve rather than a clear peak as that observed in ZnCo$_2$O$_4$. There is not a discernible difference between the cubic and tetragonal structures, so this slight adjustment to the atom locations does not have an impact on the electronic transport behavior of the overall system. Figure \ref{cu2zno4inv_properties} shows the properties of the cubic system with cation inversion. It can be seen that the materials becomes more metallic as inversion increases, and the Fermi level is brought in the midway point in the band gap; this perhaps counters the p-type dopability in these materials.

\begin{figure}
\includegraphics[width=475pt]{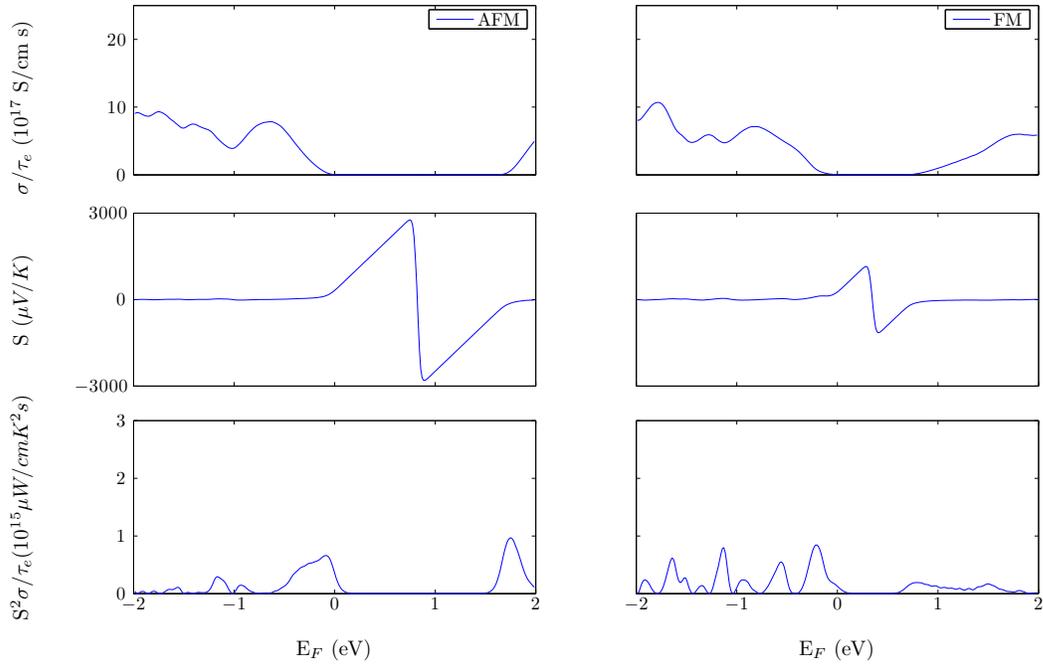}
\caption{Electronic transport properties for orthorhombic AFM and tetragonal FM spinel ZnNi$_2$O$_4$.}\label{ni2zno4_properties}
\end{figure}

Figure \ref{ni2zno4_properties} shows the electronic properties of AFM and FM tetragonal ZnNi$_2$O$_4$. The Fermi level is again located just at the edge of the valence band, and the computed intrinsic conductivities are $\sigma = 35 \ \textrm{S}/\textrm{cm}$ and $\sigma = 18 \ \textrm{S}/\textrm{cm}$ for the AFM and FM structures, respectively. In this compound we notice a less well-defined $\sigma$ local maximum and a very broad maximum range, which means that dopability for optimum performance would be easy to achieve. The conductivity maximum is found at -0.6 eV and $\sigma = 7.1 \times 10^{3} \ \textrm{S}/\textrm{cm}$ for the AFM system, and -0.8 eV and $\sigma = 7.8 \times 10^{3} \ \textrm{S}/\textrm{cm}$ for the FM system.

\section{Conclusions}
There exists a wide range of ternary oxide materials that have yet to be discovered but may exhibit valuable electrical transport and optical properties. In this work, we have studied in depth the properties of two new materials, ZnCu$_2$O$_4$ and ZnNi$_2$O$_4$, which were previously predicted to be oxide spinels \cite{Hautier2010}. We present their properties alongside those of ZnCo$_2$O$_4$, a previously studied insulator that shows promising transparent conducting properties when defect engineering principles are applied. The two new materials both show spin polarization, which could be used to fine tune their properties and switch between semiconducting and insulating behavior. As shown by theoretical computations of $\sigma$, these materials also exhibit broader conductivity maxima near the valence band edge. Hence, these materials may be easier to p-dope to induce high hole mobilities. In particular, ZnNi$_2$O$_4$ and cubic ZnCu$_2$O$_4$ exhibit high electrical conductivities while maintaining acceptable optical properties. We anticipate that these findings may guide experimental design and synthesis of new materials based on zinc oxide spinels for use in applications such as transparent conductors and magnetoelectronics.

\ack
This research was supported in part by the 3M Company through a Non-tenured Faculty Grant provided to Cynthia Lo, by the Mr. and Mrs. Spencer T. Olin Fellowship program at Washington University through a graduate fellowship provided to Maria Stoica, and by the US-India Partnership to Advance Clean Energy-Research (PACE-R) for the Solar Energy Research Institute for India and the United States (SERIIUS), funded jointly by the U.S. Department of Energy (Office of Science, Office of Basic Energy Sciences, and Energy Efficiency and Renewable Energy, Solar Energy Technology Program, under Subcontract DE-AC36-08GO28308 to the National Renewable Energy Laboratory, Golden, Colorado) and the Government of India, through the Department of Science and Technology under Subcontract IUSSTF/JCERDC-SERIIUS/2012 dated 22nd  Nov. 2012.  This work used the Extreme Science and Engineering Discovery Environment (XSEDE), which is supported by National Science Foundation grant number OCI-1053575.

\section*{References}

\end{document}